\documentclass[global,twocolumn]{svjour}
\usepackage{graphics}
\journalname{Applied Physics A}
\begin{document}
\title{X-ray magneto-optics of lanthanide materials: principles and
applications}
%\author{First author\inst{1} \and Second author\inst{2}% etc
\author{J.E.~Prieto
\and O.~Krupin \and K.~D\"obrich \and F.~Heigl \and G.~Kaindl \and K.~Starke
}                     % Do not remove
\institute{Freie Universit\"at Berlin, Arnimallee 14, D-14195 Berlin, Germany}
\date{Received: date / Revised version: date}
\maketitle
\begin{abstract}

Lanthanide metals are a particular class of magnetic materials in which
the magnetic moments are carried mainly by the localized electrons of
the 4$f$ shell. 
They are frequently found in technically relevant systems,
to achieve, e.g., high magnetic anisotropy.
Magneto-optical methods in the x-ray range are well suited to study
complex magnetic materials in an element-specific way.
In this work, we report on recent progress on the quantitative
determination of magneto-optical constants of several lanthanides in the
soft x-ray region and we show some examples of applications
of magneto-optics to hard-magnetic interfaces and exchange-coupled layered
structures containing lanthanide elements.

\end{abstract}
\section{Introduction}
\label{intro}

%Your text comes here. Separate text sections with
%\section{Section title}
%\label{sec:1}
%and \cite{RefJ}
%\subsection{Subsection title}
%\label{sec:2}
%as required. Don't forget to give each section
%and subsection a unique label (see Sect.~\ref{sec:1}).

The negligible overlap between the partially filled 4$f$ shells of 
neighboring atoms in lanthanide metals
leads to strongly localized magnetic moments, which in general contain
both an orbital and a spin part. The induced valence-band polarization
gives only a minor contribution to the magnetization, in contrast
to the predominantly {\em itinerant} moments of ferromagnetic
transition metals (TM).
The localized character of the 4$f$ moments is also responsible for the
negligibly small direct exchange interaction between lanthanide ions. Instead,
they couple only indirectly through the valence-band electrons
(RKKY interaction), which leads to ordering temperatures typically below room
temperature (RT) in lanthanide metals.
Non-vanishing orbital moments give rise to non-spherical charge distributions 
in the 4$f$ shell, which lead to strong ``single-ion" contributions 
to the magnetic anisotropy. In fact, the hardest magnetic materials known 
today are intermetallic
systems containig both RE and TM ions as in Co-Sm and Nd-Fe-B.
In these compounds, the high magnetic anisotropies are induced by
the RE ions, while the characteristic high ordering temperatures 
of the ferromagnetic TMs are retained.\cite{mhh93}

Techniques based on magneto-optical (MO) effects in the visible-light 
region are widely used for the analysis of magnetic materials.\cite{qib00}
The MO Kerr effect (MOKE) finds technological applications as well 
in the reading process of MO disks.\cite{MO}
MOKE is based on the difference in reflectivity of polarized light
upon reversal of the local magnetization direction.
Although MO effects are generally small in the visible-light region, sensitive
detection methods (lock-in techniques) yield enough contrast to allow, 
e.g., the observation of domain structures in optical microscopy.\cite{hus98}
One powerful feature of MO techniques is their capability to monitor
magnetization reversal processes in external magnetic fields.
MO experiments in the visible-light region typically involve optical
transitions between {\em delocalized} valence states. Therefore, except in
special cases, it is rather difficult to separate the magnetic
contributions of different elements.

Element selectivity has become a key issue in the analysis of magnetic
nanostructures~\cite{jfg98,mj99} or heteromagnetic systems for
information storage.\cite{MO,prince98}
It is naturally achieved by employing optical transitions that involve
core electrons, i.e. by tuning the photon energy to absorption
thresholds in the x-ray range.
The largest MO effects are found in transitions into electronic levels
of partially occupied shells which contribute to the magnetic moment.
These are the $L_{2,3}$ ($2p \rightarrow 3d$) absorption
edges of the transition metals and the $M_{4,5}$ ($3d \rightarrow 4f$)
and $N_{4,5}$ ($4d \rightarrow 4f$) thresholds of lanthanides.
All these transitions occur in the soft x-ray region.

It has been shown by several studies~\cite{cck94,tsb98,ggj01}
that in order to extract useful quantitative information from
soft x-ray MO signals in layered systems (e.g. layer-resolved
magnetization profiles), comparisons with model calculations~\cite{sts00}
of reflected intensities are required. Because the wavelength of soft x-rays
is often comparable with the dimensions of thin films and multilayers in
the nano\-meter range, one has to consider interference effects.
For their treatment in model calculations accurate values of the MO 
constants of the magnetic elements are required. While several experimental 
determinations of soft x-ray MO constants~\cite{shp98,cic98,kok00,msc00,kom01} 
and reflection coefficients~\cite{mag02} have been reported for the $L_{2,3}$ 
thresholds of TMs, results on lanthanide elements
have been scarce. Hence, we set out to experimentally determine the 
MO constants of lanthanide metals in the soft x-ray resonance regions.

The paper is organized as follows. After giving a short account of the
experimental methods used, we describe the determination of the soft x-ray MO
constants of Gd and Tb. We present then several examples of applications
of the constants as input for calculations of MO properties of thin lanthanide
films (magnetization-dependent reflectivity and Faraday rotation).
Finally, two examples of using MO effects to study epitaxial lanthanide
systems (the Sm/Co interface and Gd/Y/Tb trilayers) are discussed.

\section{Experimental}
\label{exp}

Absorption experiments at the $4d\rightarrow4f$ absorption thresholds
were performed at UE56 undulator beamlines~\cite{UE56}
of the Berliner Elektronenspeicherring f\"ur Synchrotronstrahlung
(BESSY II), while those at the $3d\rightarrow4f$ thresholds
were performed at beamline ID12-B/HELIOS-I of the European
Synchrotron Radiation Facility (ESRF).\cite{lah99,hsl99}
In the experiments at UE56, the photon energy resolution
was set to about ${\rm 100~meV}$~(full width at half maximum),
which is well below the
intrinsic width of the narrow $N_{4,5}$ pre-edge absorption
lines of Gd and Tb.\cite{sna97}
By scanning the photon energy with a synchronized movement
of monochromator and undulator, an easy normalization of the spectra
was made possible.
For the absorption measurements at the $M_{4,5}$ thresholds at ID12,
the energy resolution was set to about 0.4~eV.

Absorption spectra were recorded in total-electron yield (TEY) mode
using a high-current channeltron. To suppress the background of secondary
electrons from the chamber walls, both the sample and a retarding grid
in front of the channeltron were biased using a low-voltage battery.
For signal stability, high voltage was supplied to the channeltron 
by a 3.2-kV battery box.
The electron-yield current was amplified by an electrometer. 
All x-ray reflectivity measurements were performed at BESSY II.
The specularly reflected x-ray intensity was detected by a Si photodiode 
mounted on a rotation feedthrough inside the vacuum chamber.

Lanthanide-metal films have been prepared {\em in situ} by vapor
deposition in ultra-high vacuum on a W(110) single-crystal substrate.
External magnetic fields of up to 2~kOe were applied using a rotatable
electromagnet with a soft-iron yoke \cite{magnet} in order to
magnetize the films in-plane along the substrate
bcc[110] axis. This corresponds to the easy axis of
Gd and Tb metal films. Further details on the preparation and 
characterization of RE films can be found in Ref.~\cite{sta00}.

\section{MO constants}
\label{moconst}

\subsection{Absorption Coefficients}
\label{abscoef}

Compared to transmission methods, which give directly the absorption 
coefficients, TEY detection has the advantage to allow the use of metallic 
single crystals as substrates, on which well-characterized epitaxial films 
can be grown.
In particular, annealing of the deposited lanthanide films
at the optimum temperatures results in smooth films
with homogeneous thicknesses and high remanent magnetization.\cite{sfb92}

\begin{figure}[ht]
\begin{center}
\resizebox{0.40\textwidth}{!}{\includegraphics*{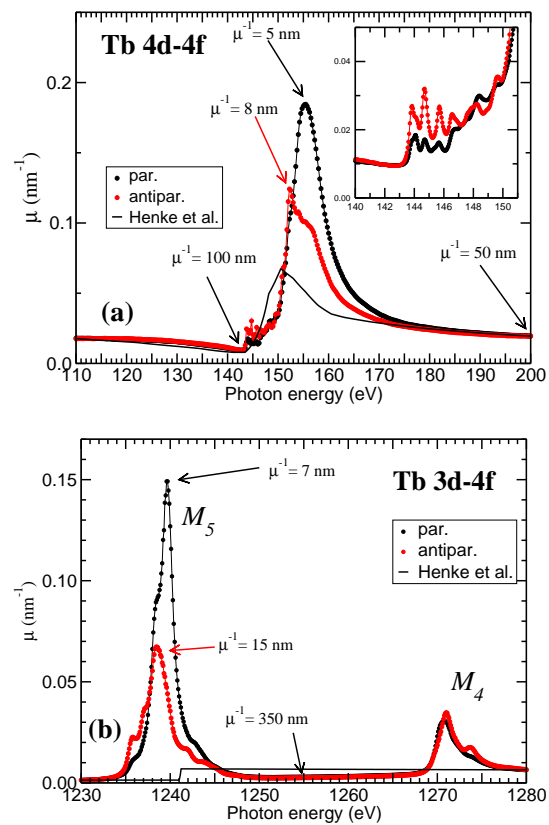}}
\caption{Absorption coefficients of Tb metal at (a) the $4d\rightarrow 4f$ 
and (b) $3d\rightarrow 4f$ resonance regions for CP light propagating
nearly parallel or antiparallel to the sample magnetization. Data
of Henke {\em et al.}~\cite{henke} are included for comparison. Values for
of the light absorption length $\mu^{-1}$ are given for selected 
photon energies.}
\label{tbmu}
\end{center}
\end{figure}

Figure~\ref{tbmu} displays experimental absorption spectra in the
Tb $N_{4,5}$ ($4d\rightarrow4f$) and $M_{4,5}$
($3d\rightarrow4f$) threshold regions with either nearly
parallel or antiparallel orientation of magnetization and
photon spin of the incoming circularly polarized (CP) light.
The spectra have been scaled to match the
tabulated values of Henke {\em et al.}\cite{henke} at the ends of the
experimental photon energy ranges after having been corrected for
intrinsic saturation. This affects the measured TEY spectra when
the attenuation length of the electromagnetic radiation, $\mu^{-1}$,
becomes comparable to $d_e$, the inelastic mean free path (IMFP) of
electrons in the solid.\cite{nsi99,vt88}
The saturation correction was performed using the relation given by
van der Laan and Thole~\cite{vt88} and has been described in detail
elsewhere~\cite{phk03}.

Lanthanide absorption spectra at the $4d\rightarrow 4f$ excitation thresholds
can be divided into two regimes, as shown in Fig.~\ref{tbmu}(a).
The so-called {\em pre-edge} lines at low energies would be
`dipole-forbidden' in the hypothetical limit of strict Russel-Saunders
coupling and their presence is only possible due to the 4$d$ spin-orbit
interaction.
Owing to small Auger matrix elements, these pre-edge lines have narrow
life-time widths of some 350 meV \cite{sna97}. In contrast, the large
absorption maxima ({\em giant resonances}) comprise several
strong and broad absorption lines. The characteristic asymmetric shape
(`Beut\-ler-Fa\-no' profile) is due to the quantum-mechanical interference 
of two excitation channels leading to the same final state: the direct 
photoemission from the 4$f$ level into the continuum and the absorption from 
the 4$d$ into the 4$f$ shell followed by a rapid super-Coster-Kronig decay.
Since both channels have similar probabilities at the lanthanide $N_{4,5}$
thresholds (as revealed by a small Fano $q$-parameter of about 3~\cite{sta00}),
the resonance shape is highly asymmetric; this in particular leads to
a strong variation of the light absorption length across the giant resonance
where values range from a few nanometers at the maxima to hundred
nanometers at the {\em antiresonance} region, as shown in 
Fig.~\ref{tbmu}(a).

At the lanthanide $M_{4,5}$ thresholds ($3d\rightarrow4f$
transitions), the photoemission channel is much weaker than the absorption
channel so that the Fano $q$ parameters are of the order of 100.\cite{sta00}
For such large $q$ values, the Fano profile approaches the Lorentzian shape 
and, in fact, the $M_{4,5}$ absorption spectra contain hundreds of
Lorentzian-shaped multiplet components.
The  spin-orbit coupling in the 3$d$ shell is the strongest interaction in
the final state, so that the components cluster into two groups,
the $M_{5}$ and $M_{4}$ thresholds,\cite{gtl88} depending on whether
the spin of the 3$d$ hole state is oriented parallel or antiparallel to the
$l$~=~2 orbital angular momentum, respectively.
As for the $N_{4,5}$ thresholds, the absorption lengths change dramatically
as the photon energy is scanned through the $M_{4,5}$ resonances 
(see Fig.~\ref{tbmu}(b)).

From the values obtained for the absorption coefficients (Fig.~\ref{tbmu}) 
we are able
to calculate the magneto-optical constants, i.e. the real and imaginary
parts of the complex index of refraction, defined as:\cite{henke}
$n_{\pm}(E) = 1 - \delta_{\pm}(E) -i \beta_{\pm}(E)$.
The $+$ and $-$ signs refer to the magnetization pointing
either parallel or antiparallel to the CP photon spin, respectively.
The imaginary part is directly related to the absorption coefficient
through
$\beta_{\pm}(E) = \frac{1}{4\pi} \frac{hc}{E} \mu_{\pm}(E)$,
while the real parts can be calculated by means of a Hilbert transformation
using the Kramers-Kronig relations.\cite{phk03}

\begin{figure}
\begin{center}
\resizebox{0.47\textwidth}{!}{\includegraphics*{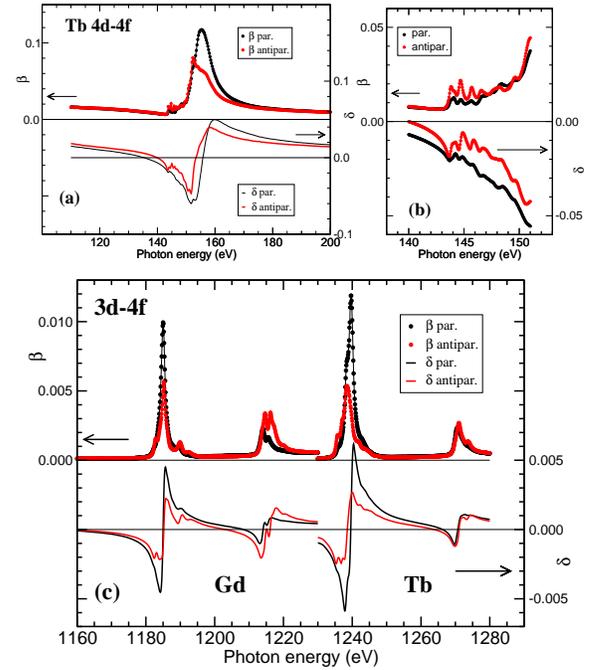}}
\caption{MO constants of (a) Tb at the $4d\rightarrow4f$;
(b) the Tb $4d\rightarrow4f$ pre-resonance structure, 
and (c) Gd and Tb at the 3d-4f threshold.}
\label{btdt}
\end{center}
\end{figure}

The calculated real parts of the refractive index of Tb at the
$N_{4,5}$ and $M_{4,5}$ absorption thresholds are shown in Fig.~\ref{btdt}
for opposite orientations of magnetization and photon spin.
The spectra exhibit the well-known dispersive behaviour with significant tails 
reaching far below and above the peaks of the associated imaginary parts. 
While the magnetic contrast of the latter 
is found mainly in the absorption maxima,
the real parts provide magnetic contrast also in regions where
the absorption is small. Since the reflected signal is determined
by both parts of the refractive index, this allows magnetization-dependent 
measurements in reflectivity with widely different pene\-tra\-tion depths of 
the incoming light.

\section{Applications}
\label{appl}

\subsection{Faraday rotation}
\label{faraday}

The MO constants can be applied in calculations of MO properties.
As a first example we consider the Faraday effect. The rotation
angle of linearly polarized (LP) light passing through a perpendicularly
magnetized film is proportional to the difference 
$\Delta n = (\delta_+ - \delta_-)$ of the real parts of the refractive
index for the two opposite light helicities into which LP light can be
decomposed. The associated difference in the imaginary parts, 
$(\beta_+ - \beta_-)$ gives rise to ellipticity, i.e. the outcoming light
is no longer linearly but elliptically polarized. The Faraday effect at
TM L$_{2,3}$ edges has been studied in Refs.\cite{kok00,msc00,kom01}.

\begin{figure}[ht]
\begin{center}
\resizebox{0.40\textwidth}{!}{\includegraphics*{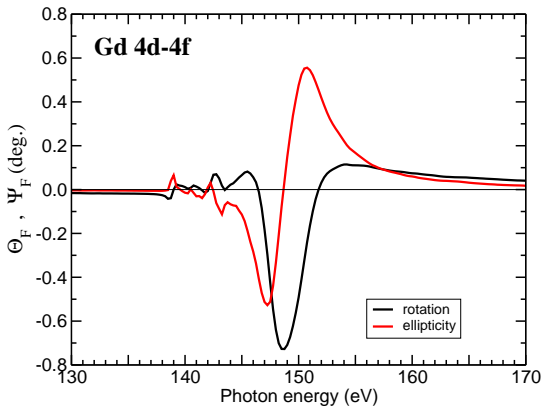}}
\caption{Calculated Faraday rotation and ellipticity for 1 ML (0.3 nm) Gd 
in the region of the $4d \rightarrow 4f$ absorption threshold.}
\label{far03}
\end{center}
\end{figure}

We have calculated the specific Faraday rotation (FR) and ellipticity for 
%a 1-monolayer (ML) thick (0.3~nm) 
a thin Gd film at the $4d \rightarrow 4f$ threshold. 
The result is shown in Fig.~\ref{far03}. 
%The large MO effects at this  
%threshold give rise to a huge specific
%Faraday rotation of about 0.7$^\circ$ per ML at 149~eV, the maximum of
%the giant resonance. 
At 149~eV, the maximum of the giant resonance, the imaginary part of the
refractive index does not 
depend on the magnetization direction, so that the ellipticity vanishes,
and the FR reaches the huge value of 0.7$^\circ$ per ML as a consequence 
of the large difference in the real parts of the refractive index 
$(\delta_+ - \delta_-)$, which in turn is due to the large difference 
in absorption $(\beta_+ - \beta_-)$.
For comparison, the specific FR at the Fe $L_{2,3}$ edge is 10 times 
smaller.\cite{kok00}
An interesting open question is whether this huge value for the FR, as 
calculated here assuming a continuous medium, will also hold for ultrathin
films of thicknesses in the range of a few monolayers.

We have proposed to use this large FR to build a ``line switch'', i.e., 
a device for fast switching between $s$ and $p$ polarization of soft x-rays 
at $h\nu=$149~eV.\cite{phk02} To this end, a 18.5~nm Gd 
film would be required in order to achieve a $90^{\circ}$ rotation of the 
light polarization plane upon reversal of the film magnetization direction. 
We expect magnetization reversal to be the time-limiting process, 
so that switching rates in the kHz-MHz region should become feasible
in this way. 
Such a device should enable highly sensitive, differential (lock-in) 
measurements.
One could extend the photon energy range of this method to about 
${\rm 180~eV}$~\cite{henke} by using the $4d \rightarrow 4f$ resonances 
of heavier lanthanide elements.
A similar scheme was proposed by Goedkoop {\em et al.} at the $M_{4,5}$
edges for the production of CP light.\cite{gft88}

\subsection{Dichroic reflectivity}
\label{dichrfl}

\begin{figure}[ht]
\begin{center}
\resizebox{0.45\textwidth}{!}{\includegraphics*{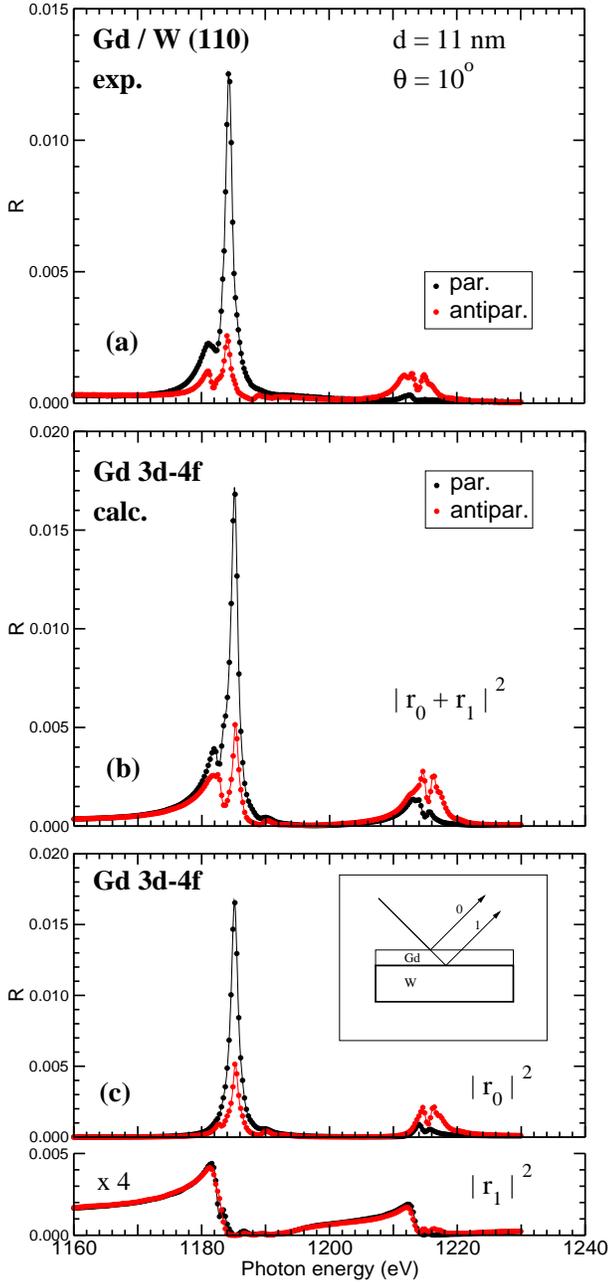}}
\caption{Experimental (a) and calculated (b) dichroic reflectivity spectra 
at the $M_{4,5}$ threshold of an epitaxial Gd/W(110) film. The film 
thickness is 11~nm and the light incidence angle 10$^\circ$. CP light was
used propagating nearly parallel or antiparallel to the in-plane sample
magnetization. Panel (c) shows the separate intensities of the two light 
paths considered in the calculations, which are schematically displayed 
in the insert.}
\label{gdrf}
\end{center}
\end{figure}

As a second application of the magneto-optical constants,
Fig.~\ref{gdrf}(b) shows calculated reflectivity spectra of
a longitudinally magnetized Gd film in the region of the $M_{4,5}$ 
threshold for CP incident light. The reflectivity was calculated using 
the Fresnel equations in the Jones matrix formalism. It includes 
longitudinal MO effects in the non-diagonal
elements which connect the $s$ and $p$ components
for transmision and reflectivity at interfaces of magnetized
media.\cite{zvk97,opp01}
We consider the reflected wave to be formed by the coherent superposition
of the amplitudes along two different light paths, as illustrated in
the insert of Fig.~\ref{gdrf}(c): (0) represents the reflection at the
vacuum/Gd interface, and (1) comprises transmission through this
interface, propagation in the Gd film (which includes absorption
and Faraday effect), reflection at the Gd/W(110) surface,
propagation back through the film, and transmission through the
Gd/vacuum interface.
Higher-order paths that include multiple reflections inside the film
are found to contribute negligibly to the reflected intensity.
In the calculation, we employed the MO constants for Gd at the
$M_{4,5}$ thresholds shown in Fig.~\ref{btdt}, together with
the values for the W substrate taken from Ref.~\cite{henke}.

Figure~\ref{gdrf}(a) shows experimental
reflectivity spectra of a Gd film on W(110). The film
thickness is $d=11\pm$1~nm and the light
incidence angle is $\theta{}=10\pm{}1^{\circ}$.
The experimental spectra are well reproduced by the calculation
shown in Fig.~\ref{gdrf}(b), including all the fine structure.
The best agreement between calculated and experimental spectra was
achieved by setting $d=12~{\rm nm}$ and $\theta{}=11^{\circ}$ in the
calculation. The experimental reflectivity R was quantified by
normalizing to the intensity of the `direct' beam, measured with the diode.
Considering the uncertainties in detector position and the
simplicity of the calculation (for example, no roughness was considered), 
the agreement of calculated with measured intensities is quite satisfactory.

It is instructive to separate the contributions of the light paths~0 and~1 
shown in Fig~\ref{gdrf}(c). While the reflected intensity
at the $M_{4,5}$ maxima is practically given by the reflectivity at the
Gd surface alone, it is in their low-energy tails where the contribution
from light paths inside the film becomes largest. Due to reduced absorption
in the flanks, the light penetration length $\mu^{-1}$ is long enough for beam~1 to return to the film surface with appreciable intensity. 
At the high-energy side of the maxima the real part of the refractive index 
($1-\delta$) is smaller than 1 [see
Fig.\ref{btdt}(c)], i.e. the light arriving from the vacuum side passes into an
optically less dense medium, so that we expect a behaviour similar to that at
the glass/air interface, where total internal reflection occurs for
incidence angles smaller than the critical angle. 
Hence, light cannot enter the Gd layer for energies right above the
absorption thresholds, while at the maxima it is greatly attenuated
inside the film.
Despite this simple analogy to visible-light optics, 
the presence of absorption
(imaginary parts of $n$) and the strong variation of the critical angle at 
the soft x-ray absorption maxima complicate the interpretation of x-ray 
reflectivity spectra.

\subsection{Sm/Co interface}
 
Co$_5$Sm is one of the hardest magnetic materials known today, 
much harder than pure Co metal.\cite{herbst91,buschow91}
Therefore it is interesting to study the effect of Sm in the magnetic 
properties of very thin Co films, eventually 
aiming at the preparation of highly anisotropic films that can retain 
high ordering temperatures and high coercivities at RT even when the
system dimensions approach the nanometer scale. We have performed a study of
the epitaxial system Sm/Co(0001) on W(110) and have found several ordered 
phases.\cite{prietoyaveremos} In the following we will concentrate on
one of these phases which appears for a Sm coverage of about 1~ML.

\begin{figure}[ht]
\begin{center}
\resizebox{0.4\textwidth}{!}{\includegraphics*{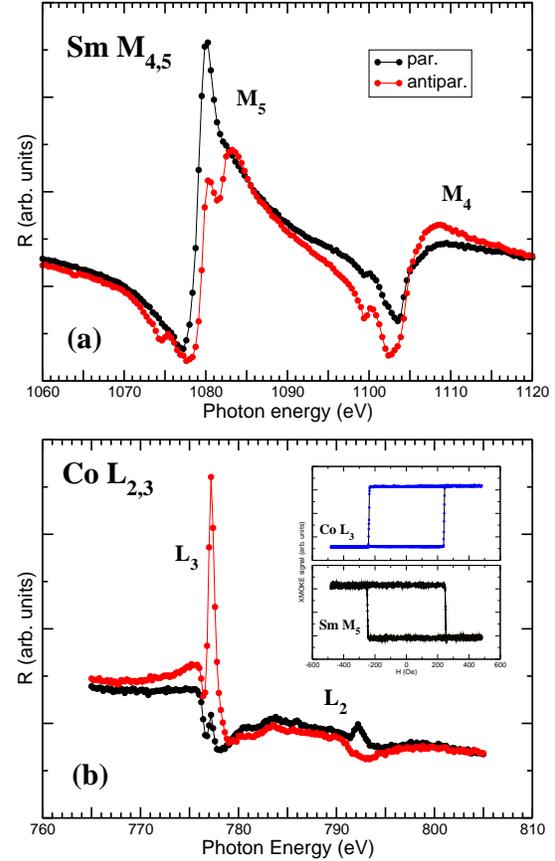}}
\caption{Dichroic reflectivity spectra at (a) the Sm $M_{4,5}$ and (b) the 
Co $L_{2,3}$ absorption thresholds of the Sm(1~ML)/Co(10~ML)/W(110) structure. 
The insert shows element-specific hysteresis loops measured at the Sm $M_5$ 
and Co $L_3$ reflectivity maxima with a light incidence angle of 10$^\circ$.}
\label{smco}
\end{center}
\end{figure}

Figure~\ref{smco} shows soft x-ray reflectivity spectra measured on a 
10~ML Co/W(110) film on top of which 1~ML Sm has been previously deposited. 
The sample was magnetized in-plane and CP light was 
used with the photon nearly parallel or antiparallel to the sample 
magnetization direction. 
Panel (a) corresponds to the region of the Sm $M_{4,5}$ threshold. Spectra 
were recorded at an incidence angle of 10$^\circ$ with respect to
the sample surface. Panel (b) shows the reflectivity spectra at the 
Co $L_{2,3}$ edge measured at an incidence angle of 20$^\circ$.
The spectra show sufficient magnetic contrast to allow the performance of
XMOKE measurements at the Sm $M_5$ and Co $L_3$ thresholds. 
The result, for an incidence angle of 10$^\circ$, is shown 
in the insert of Fig.~\ref{smco}(b). 
The magnetization reversal process takes place homogeneously over the 
illuminated area, as deduced from the square shape of
the hysteresis loops. Furthermore, the element-specific hysteresis
of both elements reveal the same coercivity $H_c$ (about 250~Oe),
showing that the film magnetization reverses simultaneously at the
Sm/Co interface and further away inside the Co film.
The present Co film (10~ML) is probably too thin to allow the observation 
of a different magnetization reversal of the two elements that would point 
to a spring-magnet behaviour.\cite{fjs98,fjg98}. 
Work is in progress aiming to prepare ordered Sm 
phases on top of thicker epitaxial Co films. 

\subsection{Gd/Y/Tb trilayers}

Exchange-coupled magnetic films through nonmagnetic spacer layers
are at the heart of many modern magnetic devices such as spin valves.
Here, the magnetic coupling is mediated by the conduction
electrons of the spacer material and its characteristics are largely
determined by their electronic structure, particularly
by the topology of the Fermi surface.\cite{bc92} The use of ferromagnetic 
rare earths (RE) as magnetic materials in exchange-coupled systems may 
become of interest, because the magnetic coupling inside the material is of
the same type (RKKY) and of comparable strength as the interlayer coupling.

\begin{figure}[ht]
\begin{center}
\resizebox{0.47\textwidth}{!}{\includegraphics*{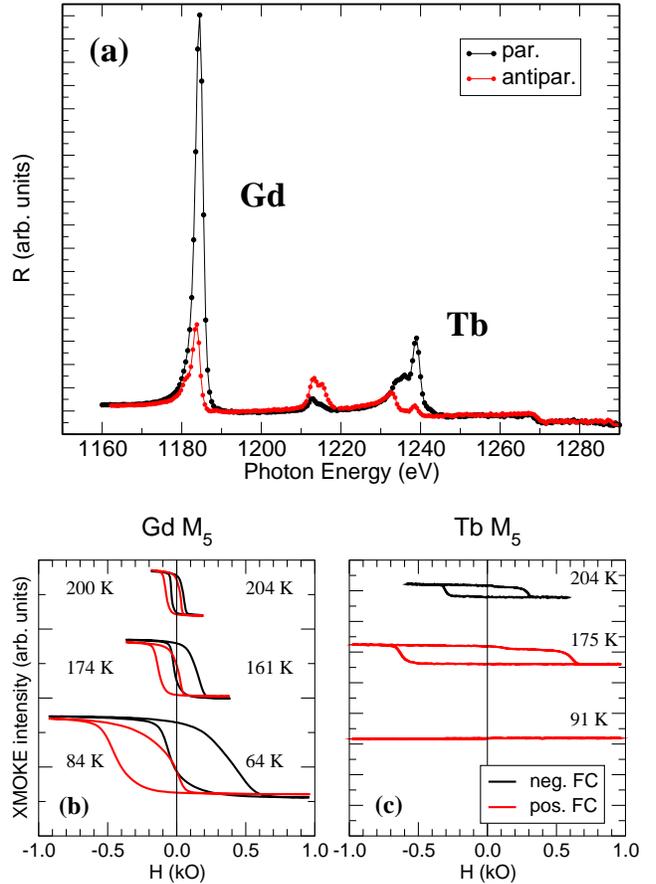}}
\caption{(a): dichroic reflectivity spectra at the $M_{4,5}$ thresholds of 
Gd and Tb in a Gd/Y/Tb trilayer structure with a 1.2~nm-thick Y spacer layer. 
The sample was cooled down to the measurement temperature (20~K) in an external field. CP light was incident at a grazing angle of 8$^\circ$ nearly parallel
or antiparallel to the sample magnetization.
Lower panel: temperature-dependent element specific XMOKE hysteresis 
curves measured at the same incidence angle with the photon energy tuned
to (b) the Gd and (c) the Tb $M_{4,5}$ reflectivity maxima. 
Black and grey lines denote the sign 
(negative and positive, repectively) of the applied field during the
cool-down of the sample to the corresponding measurement temperature.}
\label{gdytb}
\end{center}
\end{figure}

We have studied Gd~/~Y~/~Tb trilayers as a prototype
interlayer exchange-coupled RE system where the two magnetic layers
have widely different coercivities and ordering temperatures. 
While the non-spherical 4$f$ charge distribution
of Tb couples strongly to the lattice and gives highly anisotropic, 
magnetically hard films,
the ${\bf L}=0$, 4$f$ ground state of Gd (with a spherical charge 
distribution) gives much smaller coercivities. In order to ensure saturation 
magnetization of the Tb layers in our samples at low
temperatures, we cooled down the samples in external fields. 
Results of CP x-ray reflectivity measurements of an epitaxial Gd~/~Y~/~Tb 
trilayer grown on W(110) are shown in Fig.~\ref{gdytb}(a). 
The layer thicknesses are 2.0, 1.2 and 10.0~nm, respectively. 
The trilayer structure was 
magnetized in-plane along the $b$ axis (the easy axis of magnetization
of Gd and Tb films) in a field of about 2000~Oe applied 
during the cool-down to 20~K.
The photon energy range comprises the M$_{4,5}$ thresholds of
Gd and Tb. The reflectivity spectra show a sizeable MO contrast at 
the M$_5$ peak of both elements. They are ferromagnetically coupled
at this temperature.

XMOKE hysteresis curves recorded at
the M$_5$ maxima are shown in Fig.~\ref{gdytb}(b) and (c) for different 
temperatures. The black and grey lines denote the sign of the external 
field applied during the cool-down of the sample to the corresponding 
measurement temperature.
The capabibities of XMOKE as an element-specific technique are evident in this
case: the exchange field between the ferromagnetic layers Gd and Tb is
given directly by half of the horizontal shift of the Gd hysteresis recorded
after cool-down in opposite external fields [Fig.~\ref{gdytb}(b)].
For this, it is required that the magnetically hard Tb layer does not 
reverse its magnetization within the range of the applied external field, 
so that the
Gd hystereses can be considered as minor loops of the composite magnetic
structure. That this is indeed the case is shown by the Tb hysteresis 
curves in Fig.~\ref{gdytb}(c).
For 1.2~nm thickness of the Y spacer layer, the coupling between the
Gd and Tb layers is ferromagnetic and decreases strongly with increasing 
temperature, as shown by Fig.~\ref{gdytb}(b).
We attribute the main part of this effect to the 
decreasing Tb magnetization ($T_C$ of bulk Tb amounts to 220~K). 
The exchange field of 200~Oe extracted from the lowest Gd curves in 
Fig.~\ref{gdytb}(b) gives an exchange energy of 0.086 mJ/m$^2$ 
using $J=H M_s d_{Gd}$, where $M_s$ and $d_{Gd}$ are the saturation
magnetization and the thickness of the Gd layer. 
This is comparable to interlayer exchange energies found in other 
systems.\cite{chc95,pfj91}
The temperature dependence of the interlayer exchange coupling in this
RE trilayer system is under further investigation.

\section{Summary and Outlook}

By calibrating absorption spectra we have been able to quantify the
magneto-optical constants of lanthanide 
metals at the $4d\rightarrow4f$ and $3d\rightarrow4f$ excitation thresholds. 
We have demonstrated that the constants can be used
to calculate MO properties in transmission (Faraday rotation)
and reflection geometries (dichroic reflectivity spectra). Furthermore,
we have demonstrated the power of x-ray magneto-optical techniques at the examples
of a Sm/Co interface and exchange-coupled Gd/Y/Tb trilayer structures.
X-ray MO methods are expected to play an important role in 
research on magnetization dynamics in the near future, particularly 
with the advent of free-electron lasers in the soft x-ray regime.

\section{Acknowledgments}

J.E. Prieto gratefully acknowledges financial support from the
Alexander-von-Humboldt Stiftung and the Spanish MEC
(grant No. EX2001 11808094).
The authors thank F. Senf, R. Follath and H.-Ch. Mertins for experimental 
help at BESSY.
This work was financially supported by the German BMBF under Contract No. 
05 KS1 KEC/2.

%
% BibTeX users please use
\bibliographystyle{unsrt}
\bibliography{molthv2}
%X\begin{thebibliography}{}
%X\bibitem{RefJ}
%XAuthor, Journal \textbf{Volume,} (year) page numbers.
%X\bibitem{RefB}
%XAuthor, \textit{Book title} (Publisher, place year) page numbers
% etc
%X\end{thebibliography}

\end{document}